**Title: Modeling the impact of tissue oxygen profiles and oxygen depletion parameter uncertainties on biological response and therapeutic benefit of FLASH**

**Short title: Impact of tissue oxygen profile on FLASH**

Hongyu Zhu[1], Jan Schuemann[2], Qixian Zhang[3], Leo E Gerweck[2*]

1. Department of Radiation Oncology, Sun Yat-sen University Cancer Center, State Key Laboratory of Oncology in South China, Collaborative Innovation Center for Cancer Medicine, Guangzhou 510060, China
2. Department of Radiation Oncology, Massachusetts General Hospital, Boston, MA 02114, United States of America
3. Department of Radiation Oncology, Fudan University Shanghai Cancer Center, Shanghai 200032, China

**Correspondence**:

Leo E Gerweck, 55 Fruit St, Cox 7, Boston, MA 02114, United States of America, lgerweck@mgh.harvard.edu

**Funding information**

This work was funded by China Postdoctoral Science Foundation (2021M703671, H. Zhu), the Science and Technology Planning Project of Guangzhou (202201011076, H. Zhu) the National Institutes of Health/National Cancer Institute (NIH/NCI grant no. R21 CA252562, J. Schuemann), the Damon Runyon-Rachleff Innovation Award 57-19 and the Brain Tumour Charity Award GN-000642 (J. Schuemann and L. Gerweck).

**Conflict of Interest Statement:** None.

**Abstract**

**Background:** Ultra-high dose rate (FLASH) radiation has been reported to efficiently suppress tumor growth while sparing normal tissue, however, the mechanism of the differential tissue sparing effect is still not known. Oxygen has long been known to profoundly impact radiobiological responses, and radiolytic oxygen depletion has been considered to be a possible cause or contributor to the FLASH phenomenon.

**Purpose:** This work investigates the impact of tissue $pO_2$ profiles, oxygen depletion per unit dose ($g$), and the oxygen concentration yielding half-maximum radiosensitization (the average of its maximum value and one) ($k$) in tumor and normal tissue.

**Methods:** We developed a model that considers the dependent relationship between oxygen depletion and change of radiosensitivity by FLASH irradiation. The model assumed that FLASH irradiation depletes intracellular oxygen more rapidly than it diffuses into the cell from the extracellular environment. Cell survival was calculated based on the linear quadratic-linear model and the radiosensitivity related parameters were adjusted in 1 Gy increments of the administered dose. The model reproduced published experimental data that were obtained with different cell lines and oxygen concentrations, and was used to analyze the impact of parameter uncertainties on the radiobiological responses. This study expands the oxygen depletion analysis of FLASH to normal human tissue and tumor based on clinically determined aggregate and individual patient $pO_2$ profiles.

**Results:** The results show that the $pO_2$ profile is the most essential factor that affects biological response and analyses based on the median $pO_2$ rather than the full $pO_2$ profile can be unreliable and misleading. Additionally, the presence of a small fraction of cells on the threshold of radiobiologic hypoxia substantially alters biological response due to FLASH oxygen depletion. We found that an increment in the $k$ value is generally more protective of tumor than normal tissue due to a higher frequency of lower $pO_2$ values in tumors. Variation in the $g$ value affects the dose at which oxygen depletion impacts response, but does not alter the dose dependent response trends, if the $g$ value is identical in both tumor and normal tissue.

**Conclusions:** The therapeutic efficacy of FLASH oxygen depletion is likely patient and tissue dependent. For breast cancer, FLASH is beneficial in a minority of cases, however, in a subset of well oxygenated tumors, a therapeutic gain may be realized due to induced normal tissue hypoxia.

## 1 Introduction

In 2014, Favaudon et al. reported that a single dose of 20 Gy radiation with electrons administered to the thorax of rats at a mean dose rate ≥ 40 Gy/s (FLASH irradiation) resulted in "no lung complications", whereas 15 Gy administered at a conventional (CONV) dose rate (≤ 0.03 Gy/s) lead to significant lung fibrosis [1]. Additionally, FLASH and CONV irradiation equally suppressed tumor growth [1-3]. The sparing of normal rather than tumor tissue has given rise to significant clinical interest in extremely high dose-rate radiation for the treatment of cancer as well as investigations into the potential mechanism of the FLASH effect. Although several hypotheses have been advanced to explain the normal tissue sparing effect of FLASH irradiation, including rapid radiolytic oxygen depletion, changes in the production and processing of reactive oxygen species, immune response, microenvironmental factors, and others, the mechanism(s) of tissue sparing have not yet been resolved [4-12]. Additionally, most but not all studies have reported normal tissue sparing at FLASH dose rates [13,14], and Adrian et al reported a FLASH effect in several tumor cell lines in vitro [15].

It has long been known that both CONV and FLASH irradiation deplete dissolved oxygen in aqueous solutions. The principal mechanism of oxygen depletion, i.e., the binding of oxygen with primary or secondary radical products of water radiolysis suggests that radiation induced hypoxia is likely to occur in cells and tissues if the rate of intracellular oxygen depletion exceeds the rate of oxygen resupply from the extracellular environment, regardless of whether the cells and tissues are normal or malignant. Given the pronounced radiosensitizing effect of oxygen, the therapeutic efficacy of FLASH irradiation will thus likely be impacted by radiation induced hypoxia in tumor and normal tissue. It is of note that the earliest and perhaps most frequent validation of FLASH tissue sparing has been reported for skin, which is known to be on the threshold of radiobiologic hypoxia in its normal state (5-10 mmHg oxygen) [8,16-20].

Biologic effects of radiation principally arise from damage to DNA. This damage may result from direct interactions of ionizing radiation with DNA, or indirectly from the interaction between chemical products generated by the radiolysis of water and DNA. Most indirect lethal damage is caused by the hydroxyl radical OH$^\bullet$. The resulting DNA$^\bullet$ radical may either be restored to its undamaged state by hydrogen donation, primarily by amino thiols such as glutathione, cysteine, and cysteamine, or oxidized by oxygen, leading to the formation of peroxides which "fix" the DNA damage, i.e., make the damage permanent [21-23]. The fate of the DNA$^\bullet$ radical is thus dependent on competition between oxygen for damage fixation, and thiols for damage repair. The oxygen-thiol competition model for fixation or restoration of the DNA radical as well as competing radiochemical processes in mammalian cells has been validated and summarized by Koch [22]. In both bacteria and mammalian cells, the oxygen concentration needed to achieve the average of its maximum value and one (commonly referred to as half-maximum) sensitization, which is usually denoted as $k$, is increased in the presence of added thiols and decreased upon thiol depletion [22,24].

At sufficiently high doses and dose-rates, when the rate of cellular oxygen depletion exceeds the rate of oxygen diffusion into cells, both bacteria and mammalian cells exhibit a pronounced decrease in sensitivity to radiation, and the dose at which the sensitivity to radiation decreases is directly dependent on the initial oxygen concentration [25-28]. While these quantitative studies and results have largely been pioneered and demonstrated in bacteria and mammalian cells in vitro, the impact of radiobiologic hypoxia on the response of tumors, normal tissues and spheroids also yield oxygen enhancement ratios (OER) of approximately 2.5-3.0 [29-32]. In short, small naturally occurring or induced changes in oxygen status may significantly impact cell and tissue response to irradiation. This effect becomes especially significant in the context of stereotactic body and FLASH irradiation, which utilize doses in the range of 10-20 Gy per fraction.

The extent to which FLASH oxygen depletion impacts tissue response will depend on pretreatment tissue pO$_2$, oxygen depletion per unit dose, total dose and the oxygen concentration at which half-maximum sensitization occurs. In this study, we examined two descriptors of tissue oxygenation, i.e., median tissue pO$_2$ and complete pO$_2$ tissue profiles, and the impact of the reported ranges and uncertainties in the aforementioned parameters on FLASH oxygen depletion and the resultant change in cell response. The results show that the therapeutic efficacy of FLASH is likely patient dependent. We identify circumstances under which FLASH oxygen depletion could be of therapeutic benefit or deficit.

## 2 Methods

### 2.1 Modeling the impact of FLASH oxygen depletion on cellular response

Based on decades of evidence, this study assumes that the oxygen concentration of tumor and normal tissue is a determinant of response to radiation. To evaluate the potential impact of oxygen depletion, including uncertainties in the oxygen depletion ($g$) per unit dose and the oxygen concentration at which the OER reaches the average of its maximum value and one ($k$), cell surviving fractions (SF) were calculated based on the linear quadratic-linear (LQ-L) model [33,34]:

$$\ln(\text{SF}) = -(\alpha D + \beta D^2), \qquad \text{for } D \leq D_T \tag{1a}$$

$$\ln(\text{SF}) = -\left(\alpha D_T + \beta D_T{}^2\right) - \gamma(D - D_T), \qquad \text{for } D > D_T \tag{1b}$$

$\alpha$ and $\beta$ are inactivation parameters which characterize cell and tissue response to radiation, $D_T$ is the transition point at which the SF curve becomes linear, $\gamma = -(\alpha + 2\beta D_T)$ is the log cell kill per Gy in the linear portion of the logarithmic survival curve as determined by the slope of the line tangent to the LQ curve at dose $D_T$. The choice of an LQ-L model to characterize the response of cells and tissues to radiation, and oxygen modification of response, is arbitrary. Response trends due to FLASH oxygen depletion vs. conventional dose rate radiation do not differ when characterized by the LQ and LQ-L model.

To estimate the impact of oxygen on cell response, the method proposed in the reference was used to modify the parameters of $\alpha$ and $\beta$ [35]. i.e.,

$$\alpha_{\text{aerobic}} = \alpha_{\text{anoxic}} \times \text{OER} \tag{2}$$

$$(\alpha/\beta)_{\text{aerobic}} = (\alpha/\beta)_{\text{anoxic}}/\text{OER} \tag{3}$$

$$\text{OER} = \frac{k + m \times [O_2]}{k + [O_2]} \tag{4}$$

OER was calculated with the empirical function proposed by Alper and Howard-Flanders [36]. $m$ is the maximum OER and $k$ is the oxygen concentration (mmHg) at which the OER is equal to the average of its maximum value and one. $[O_2]$ is the oxygen concentration (mmHg). In this study, $m$ was assumed to be 3. The transition point $D_T$ for cells with different oxygen concentration in the LQ-L model was calculated with:

$$D_T|_{\text{aerobic}} = D_T|_{\text{anoxic}}/\text{OER} \tag{5}$$

## 2.2 Model evaluation and processing of tissue pO₂ profiles

To evaluate the validity of the model we determined whether it predicted the experimental results reported by Ling et al. [37] and Michaels et al. [38], as shown in Figure 1. These investigators placed attached CHO cells coated with a thin film of medium into a humidified 100% N₂ environment or an environment containing 0%, 0.21% and 0.44% oxygen in N₂. The cells were then exposed to single dose irradiation of 3 ns duration. At a dose which deleted all oxygen, the cell sensitivity was identical to the sensitivity of cells irradiated under 100% N₂ conditions. To quantify the impact of oxygen depletion on the radiation sensitivity parameters $\alpha$, $\beta$ and $\gamma$, we utilized the Alper and Howard-Flanders competition model [36] and updated the cell surviving fraction for 1 Gy fractions of the total dose. Practically, this was implemented via the following steps:

1. Deliver the $n^{\text{th}}$ fractional dose $\Delta D$ and decrease the intracellular oxygen concentration by the value of $g \times \Delta D$; and the total delivered dose $D_n = n \times \Delta D$.

2. Calculate the $\text{OER}_n$, and the $\text{OER}_n$ adjusted $\alpha_n$, $\beta_n$, $\gamma_n$, and $D_{Tn}$ according to Eq.1-5, and calculate $SF_n(D_{n-1})$ and $SF_n(D_n)$ according to Eq.1 with the updated parameters for this n-th fractional dose.

3. Calculate the fractional decrease in the surviving fraction $d_{SF|n} = \frac{SF_n(D_n)}{SF_n(D_{n-1})}$. The SF after the $n^{\text{th}}$ fractional FLASH dose is then calculated in a recursive manner for each fractional dose: $SF_{FLASH}(D_n) = SF_{FLASH}(D_{n-1}) \times d_{SF|n}$

The same oxygen depletion and radiation survival model (LQ-L) was applied to more complex tissue pO₂ profiles, i.e., containing well oxygenated foci as well as low pO₂ foci in the same tissue. Briefly, the percent cells or foci within a pO₂ range such as 0-2.5, 2.5-5, 5-7.5 mmHg up to the highest recorded pO₂ value is processed in the same way as described for cells. All cells in each bin are assumed to be at the same $pO_2$ i.e., the mid $pO_2$ value of each bin e.g., 1.25 mmHg in the 0-2.5 mmHg bin. The sum of the surviving fractions in each 2.5 mmHg bin was then calculated. As each 1 Gy fractional dose depletes oxygen, alpha and beta values are accordingly recalculated for each additional 1 Gy fractional dose.

## 2.3 Application of the model to human tissue; methodology and assumptions

A treatment site (breast) for which substantial $pO_2$ data is available for both tumor and normal tissue [39] was selected to evaluate the effects of FLASH vs. CONV irradiation in human tissue with a heterogeneous $pO_2$ distribution. Vaupel et al. [39] obtained aggregate normal breast $pO_2$ profiles of $N$=16 patients, $n$=1009 evaluated foci along with breast tumor $pO_2$ profiles in 15 of the same $N$=16 patients, $n$=1068 foci, and the $pO_2$ profile of two individual patient's breast tumor assessed by the Eppendorf polarographic system. The $pO_2$ profiles of normal human brain and subcutis were also extracted from Vaupel et al. for analysis [40]. The $pO_2$ profiles were extracted using the GetData graph digitizer (http://getdata-graph-digitizer.com/), and presented as the relative frequency of tissue in each 2.5 mmHg $pO_2$ bin (Figure 2).

The SF responses of normal breast and breast tumor were calculated by the following method:

1. For CONV irradiation, it was assumed that the oxygen supply exceeded the rate of oxygen depletion and the tissue pO2 profile was unchanged.

2. For FLASH irradiation, the $pO_2$ profile was shifted by $g \times \Delta D$ after each $\Delta D$ =1 Gy fractional dose; the parameters OER, and OER adjusted $\alpha, \beta, \gamma, D_T$ values were updated in each bin to calculate the SF using the fractional dose method described above.

3. Radiolytic oxygen depletion is regional and equally applies to both the cellular and extracellular compartments. It is assumed that intracellular and extracellular oxygen depletion exceeds the rate of oxygen resupply from the nearest oxygen rich precapillary arterioles and capillaries [41].

## 2.4 Parameter values in the model

To reproduce the experimental data reported by Ling et al. [37] and Michaels et al. [38] the parameter values $\alpha_{\text{anoxic}}$ = 0.0156 Gy$^{-1}$ and $\beta_{\text{anoxic}}$ =0.0071 Gy$^{-2}$ were determined by fitting the SF data measured under anoxic condition from the experimental $N_2$ SF curve with the LQ-L model. $g$ =

0.275 mmHg/Gy was adopted as their data indicate 12 Gy depletes 0.44% oxygen (3.3 mmHg), similar to the average $g$ value reported for mammalian cells, (Supplemental material section 1). The estimated 12 Gy transition dose $D_T$ was determined based on the experimental SF curve shape, i.e., the point at which the dose response curve became linear as reported by Michaels et al. [38].

For the analysis of normal human breast and breast tumor tissues, the $\alpha$ and $\beta$ values for breast tumor were adopted from [42] with $\alpha = 0.374$ Gy$^{-1}$, $\beta = 0.0251$ Gy$^{-2}$. Late skin response ($\alpha = 0.0432$ Gy$^{-1}$, $\beta = 0.0227$ Gy$^{-2}$) was considered as a surrogate for normal breast [43]. $D_T|_{aerobic}$ was set as 10 Gy for both normal breast and breast tumor.

Differences in the intracellular concentration of aminothiols between tissue types as well as in vitro and in vivo, have been reported and shown to impact the value of $k$. $k = 3.8$ mmHg was used for the calculation of cell SF, while the range of 3.8-15 mmHg was considered to evaluate the impact of the uncertainty of $k$ on cell survival [22-24,41,44].

The parameter $g$, that is oxygen depletion per unit dose, is one of the most impactful factors in FLASH oxygen depletion. In this work we evaluated the minimum and maximum reported values of $g$, i.e., $g = 0.19$ and 0.71 μM/Gy (0.15 and 0.56 mmHg/Gy) to investigate the impact of the $g$ value uncertainty on cell survival. A detailed summary of previously reported value of g can be found in the supplementary material section 1. We further used the mean reported $g$ (0.45 μM/Gy, i.e., 0.36 mmHg/Gy) as the FLASH oxygen depletion rate for other calculations.

For determination of the 95% confidence interval (CI) of the number of recorded $pO_2$ values in each 2.5 mmHg bin, the percent observations in each bin were multiplied by the total number of observations for all bins, a detailed method [45] to calculate the 95% CI is provided in the supplementary material section 2.

## 3 Results

### 3.1 Evaluation and validation of the oxygen depletion and LQ-L models

Figure 1, panels a and b show the results of in vitro studies of Ling et al, 1978 and Michaels et al 1978. Cells were equilibrated with a gas phase environment of 100% $N_2$, 0.21% and 0.44% $O_2$. The cells were exposed to doses of radiation as previously described. The surviving fractions of cells irradiated under $N_2$, 0.21% and 0.44% $O_2$ conditions are indicated by the symbols. The oxygen depletion and LQ-L predicted surviving fraction results are indicated by the dashed curves. Figure 1a also shows that a factional dose of 1 Gy or 0.1 Gy yields similar results, indicating that the 1 Gy factional dose is sufficient to describe the oxygen depletion process and cell kill which occur over the same fractional dose scale. Essentially identical results were obtained in an examination of the predicted vs. observed results in HeLa cells (supplementary Figure S1).

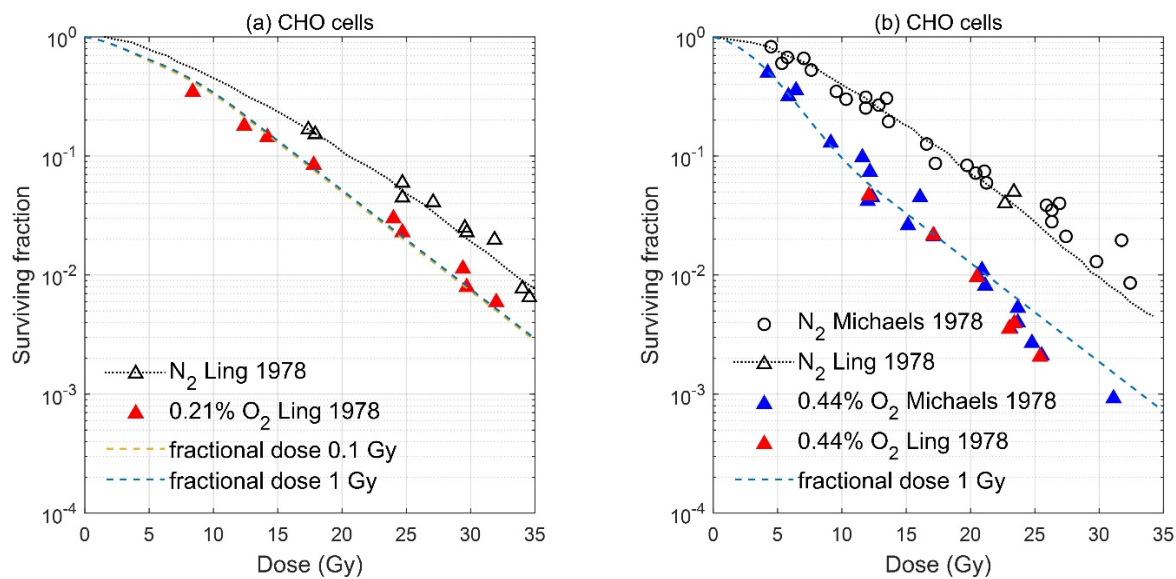

Figure 1. The surviving fraction of Chinese hamster ovary (CHO) cells after 3 ns electron FLASH irradiation under different initial oxygen concentrations. The data of cells equilibrated with nitrogen is indicated by the black dotted line and open circles as well as open triangles; 0.21% and 0.44% oxygen by blue and red triangles. Experimental data are extracted from Ling et al. [37] and Michaels et al. [38]. The colored dash lines represent the calculated results obtained with the fractional dose method for $\alpha_{anoxic}$ = 0.021 Gy$^{-1}$, $\beta_{anoxic}$ = 0.0071 Gy$^{-2}$ and $D_T$ = 12 Gy.

## 3.2 Quantifying tissue oxygenation: median pO₂ values are not appropriate for predicting the impact of FLASH irradiation.

Tissue oxygenation is frequently characterized by a single parameter such as median pO$_2$. However, modulation of radiation sensitivity occurs over a narrow pO$_2$ range, i.e., < 15 mmHg, and most prominently over the 1 to 7 mmHg range. Median or mean pO$_2$ does not reflect the percent of tissue in the 0-15 mmHg range, and more specifically, the fraction of cells in the 0-2.5 mmHg range, 2.5-5, 5-7.5 mmHg range, etc. Reported FLASH oxygen depletion values (0.19 - 0.71 μM/Gy) suggest that FLASH irradiation may reduce the $pO_2$ of tissue on the threshold of radiobiologic hypoxia, into a substantially radiation protected $pO_2$ environment. Based on median $pO_2$ values, this induced hypoxic radioprotection would generally be thought to be highly unlikely.

Figure 2 compares the response of normal human brain, subcutis [40] and breast cancer [39] to CONV vs. FLASH dose rate irradiation for median $pO_2$ values (red lines) and full $pO_2$ profiles (black lines). The most prominent features are that median tissue $pO_2$ value does not reflect the presence and frequency of low $pO_2$ values (e.g., compare the fraction of low $pO_2$ values in breast cancer, median $pO_2$ = 30 mmHg, with the low $pO_2$ fraction in normal brain median $pO_2$ = 24 mmHg). Similarly, median $pO_2$ values do not reveal the large response difference of tissue which contain a large fraction of hypoxic cells (breast cancer) when exposed to CONV vs. FLASH irradiation (panel c). Neither do median $pO_2$ values suggest that tissue such as normal brain may benefit from FLASH vs. CONV irradiation, or that the response of breast tumor which has a relatively high median $pO_2$ and large percentage of cells at very low $pO_2$, exhibits a differential response when exposed to FLASH vs. CONV dose rate irradiation. To summarize, median or average tissue $pO_2$ is an unreliable and likely misleading parameter of tissue response to FLASH irradiation.

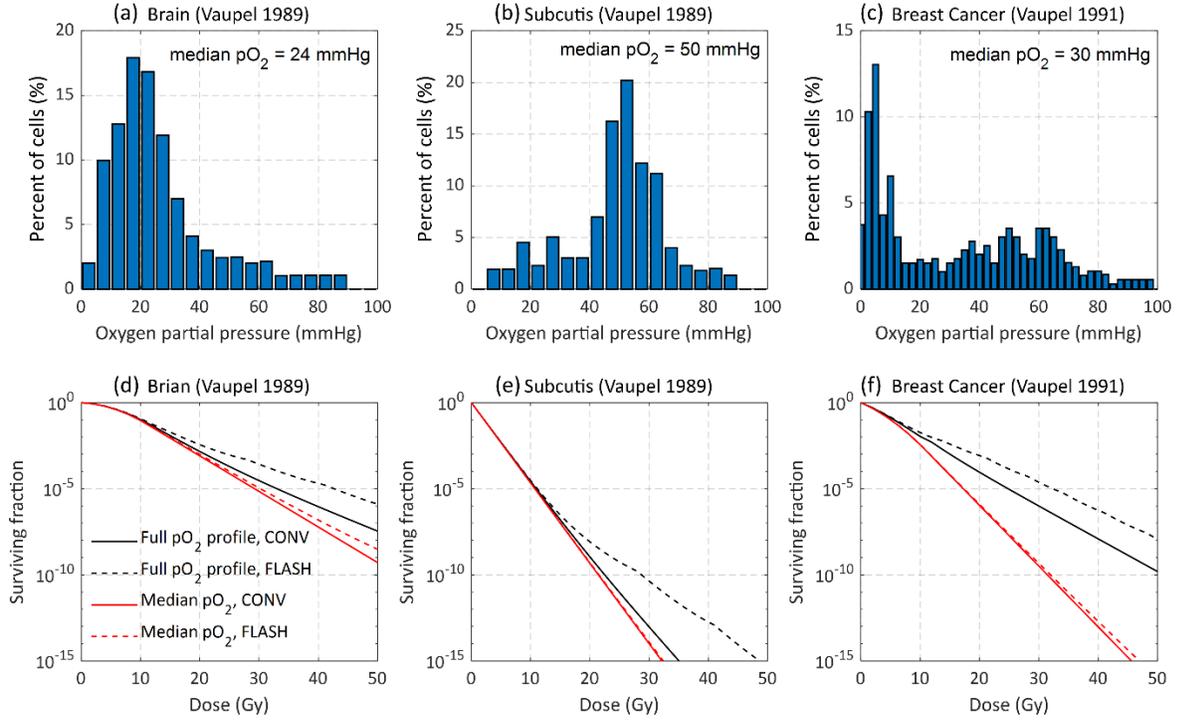

Figure 2. The pO$_2$ profile of (a) brain, (b) subcutis, and (c) breast cancer, data adopted from references [39,40]] and the (d-e) SF curves calculated with the full pO$_2$ profile or median pO$_2$. α$_{Brain}$ = 0.0499 Gy$^{-1}$, β$_{Brain}$ = 0.0238 Gy$^{-2}$ [43]; α$_{Subcutis}$ = 1.13 Gy$^{-1}$, β$_{Subcutis}$ = 0.0 Gy$^{-2}$ [46] ; α$_{Breast\ Cancer}$ = 0.374 Gy$^{-1}$, β$_{Breast\ Cancer}$ = 0.0251 Gy$^{-2}$ [42]

### 3.3 Impact of the pO$_2$ profile uncertainty

Figure 3 shows the pO$_2$ profile of aggregate normal breast, aggregate breast tumor and patient specific breast tumors before and after 20 Gy FLASH irradiation, along with the 95% CI of the percent of cells in each bin. The final frequency of tissue in each 2.5 mmHg bin was obtained by following 3 steps after each 1 Gy fractional dose: ① shift the pO$_2$ profile by $g$ × fractional dose; ② calculate the surviving fraction in each bin using the adjusted $α$, $β$, $γ$, $D_T$ values; ③ repeat the process until a preset total dose is delivered; ④ normalize the pO$_2$ profile so that the sum of frequencies in all bins equals 100% (A graphic description is provided in supplementary Figure S2).

Aggregate breast tumor, panel b, blue line, is substantially more hypoxic than normal breast prior to irradiation, panel a blue line, with approximately 15% of pO$_2$ values being < 5 mmHg, and 30%

< 10 mmHg. Twenty Gy FLASH irradiation further reduces the $pO_2$ by approximately 7.2 mmHg and 99.36 % of cells surviving 20 Gy reside in the 0-2.5 mmHg bin as indicated by the orange curve. However, substantial intertumoral $pO_2$ heterogeneity is also suggested by the approximately 50% of $pO_2$ values greater than 20 mmHg. This is seen in the breast cancer $pO_2$ profiles of patients A and B, Figure 3c and 3d. None of the tumor cells of patient A exhibits radiobiologic hypoxia after 20 Gy FLASH irradiation. In contrast, in patient B, approximately 16% of all cells are between 0 and 2.5 mmHg, and approximately 50% of cells' $pO_2$ values are < 5 mmHg prior to irradiation. Following 20 Gy irradiation, 99.85% of all surviving cells reside in the 0-2.5 mmHg $pO_2$ bin category. The very significant increase in the fraction of surviving cells in the 0-2.5 mmHg bin following 20 Gy FLASH irradiation, is due their greater radiation resistance and to the shift of cells at higher $pO_2$ to the lower $pO_2$ bin due to oxygen depletion. In aggregate normal breast and breast cancer of patient A, 20 Gy FLASH oxygen depletion is insufficient to reduce the $pO_2$ values of cell population below approximately 7.5 mmHg in normal breast and 35 mmHg in patient A breast cancer. There is very little change in the sensitivity of normal breast, and no change in the sensitivity of patient A cell due to FLASH oxygen depletion.

The profile after 20 Gy CONV irradiation is shown in supplementary Figure S3. In the absence of oxygen depletion, the percent of surviving cells following 20 Gy CONV radiation in aggregate breast and patient B breast cancer in the 0-2.5 mmHg bin, is lower than following FLASH irradiation and the overall surviving fraction is reduced.

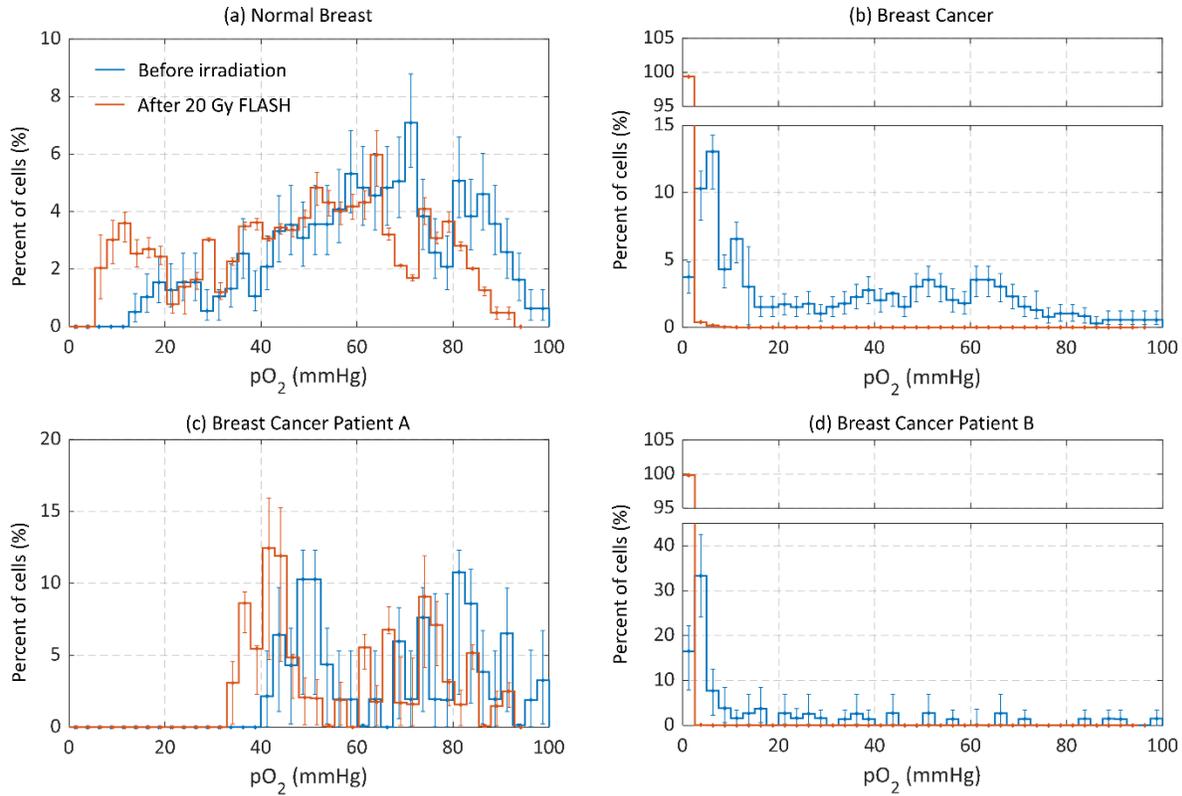

Figure 3. The pO$_2$ profile of aggregate normal breast, aggregate breast tumor and individual patient breast tumor profiles prior to and following 20 Gy FLASH irradiation. The pO$_2$ frequency distribution prior to irradiation for all cells is indicated by the blue curve, and the profiles of surviving cells following 20 Gy is indicated by the orange curve. The 95% confidence interval of the frequency of cells in each bin is shown.

### 3.4 Impact of the value of *k*

Figure 4 shows the impact of the value of *k*. With increasing *k*, cells exhibit hypoxic resistance at higher oxygen concentrations independent of dose-rate. In aggregate breast normal tissue, the value of *k* negligibly impacts the response to FLASH vs. CONV irradiation at doses less than 25 Gy. Again, this is due the relatively high minimum oxygen concentration values observed in normal breast and the low frequency of these lower oxygen concentration values. For aggregate breast cancer, an increase in the value of *k* is apparent at lower doses due to the relatively hypoxic status of breast cancer. However, the general trend does not apply to well oxygenated breast cancers such as seen in Figure 3, Patient A. In this case, an increase in $k$ may result in normal tissue protection.

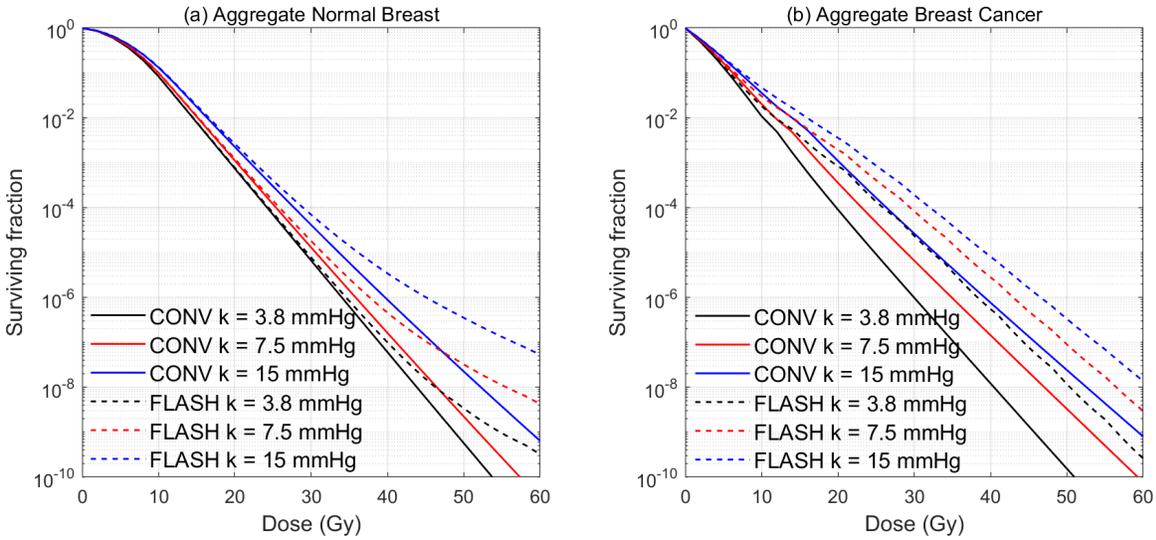

Figure 4. The impact of $k$ on the surviving fractions of (a) normal breast and (b) breast cancer. g = 0.45 µM/Gy (0.36 mmHg/Gy).

### 3.5 Impact of the value of $g$

Figure 5 shows that the impact of oxygen depletion rate ($g$). Increasing the depletion rate from 0.19 to 0.71 µM/Gy does not alter the response to CONV dose-rate irradiation as the rate of oxygen diffusion into cells likely exceeds oxygen depletion. Due to the absence of pO$_2$ values below 12.5 mmHg in normal tissue and the paucity of values in the 12.5-15 mmHg range, the impact of oxygen depletion is not apparent for values of $g$ from 0.19 to 0.71 µM/Gy below a dose of 25 Gy. Similarly, the impact of $g$ = 0.19 µM/Gy increases the dose at which FLASH induced tumor hypoxia is apparent to >10 Gy.

One may notice that the SF curves in Figure 4b and 5b are not perfectly smooth. This results from the changing $D_T$ for cells in different pO$_2$ bins. The effect of changing OER and its associated $\alpha$, $\beta$, $\gamma$, and $D_T$ is more pronounced at lower oxygen concentrations. A substantial percent of tumor tissue resides in the low pO$_2$ region and the values in each pO$_2$ bin are very different, resulting in the wavy curve. This trend is not seen in Figures 4a and 5a as most normal breast tissue exhibits higher oxygen concentrations which do not impact the parameters $\alpha$, $\beta$, $\gamma$, and $D_T$.

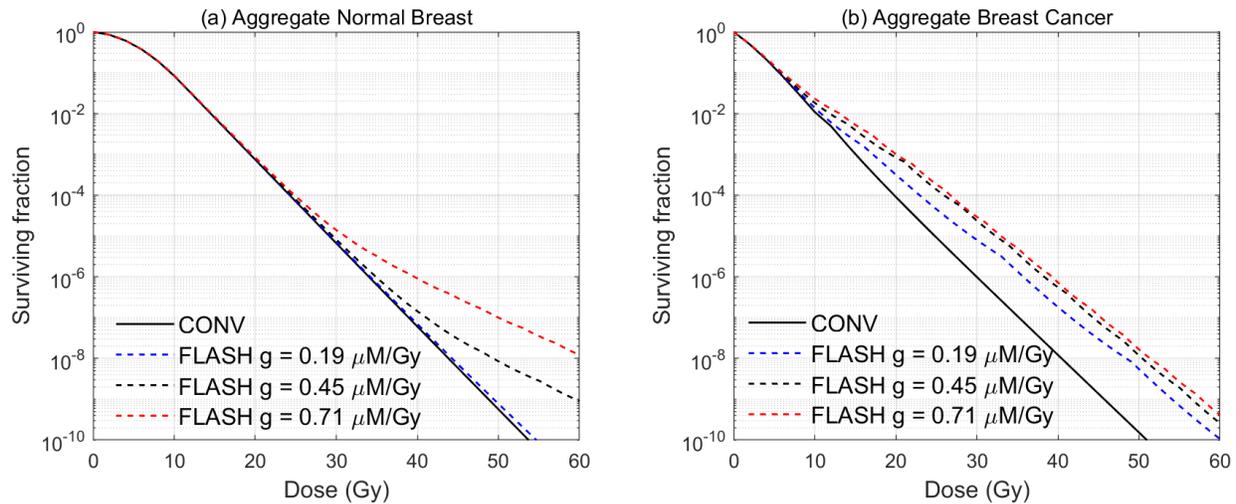

Figure 5. The impact of the value of $g$ on the surviving fractions of (a) aggregate normal breast and (b) aggregate breast cancer. $k = 3.8$ mmHg.

## 3.6 Patient therapeutic benefit

Figure 6 shows SF versus dose curves for aggregate normal breast and breast tumors in panels a and b, and individual patient tumors in panels c and d. The SF curves show that FLASH becomes protective of normal breast tissue due to induced hypoxia starting at doses of approximately 30 Gy. The sparing effect of FLASH induced hypoxia is apparent at a significantly lower dose, i.e., at approximately 10 Gy in aggregate breast tumors. Thus, for the population average $pO_2$ profile, FLASH might be expected to have a negative therapeutic effect relative to CONV irradiation.

However, aggregate cancer patient response does not predict individual patient response. For patient A, FLASH oxygen depletion by a 30 Gy dose is insufficient to reduce any of the measured tumor $pO_2$ values lower than 35 mmHg or impact tumor response. In contrast, FLASH increases tumor hypoxia at doses exceeding 10 Gy in patient B. Although FLASH and CONV dose rate irradiation yield essentially identical survival curves the existence of a small fraction of marginally radiobiologically hypoxic cells in normal tissue can substantially reduce the dose at which a radioprotective effect is observed. This is illustrated by the analysis of the impact of the addition of 1% of cells to the 2.5–5 mmHg $pO_2$ bin to the aggregate normal breast tissue profile, supplemental Figure S4. In contrast to the impact of 1% cells on the fraction of cells following CONV irradiation, exposure to FLASH irradiation increases the SF by a factor of 25.12 at 20 Gy. Figure 6 also illustrates the modest effect that the pO$_2$ frequency uncertainty per bin (95% confidence interval) has on response.

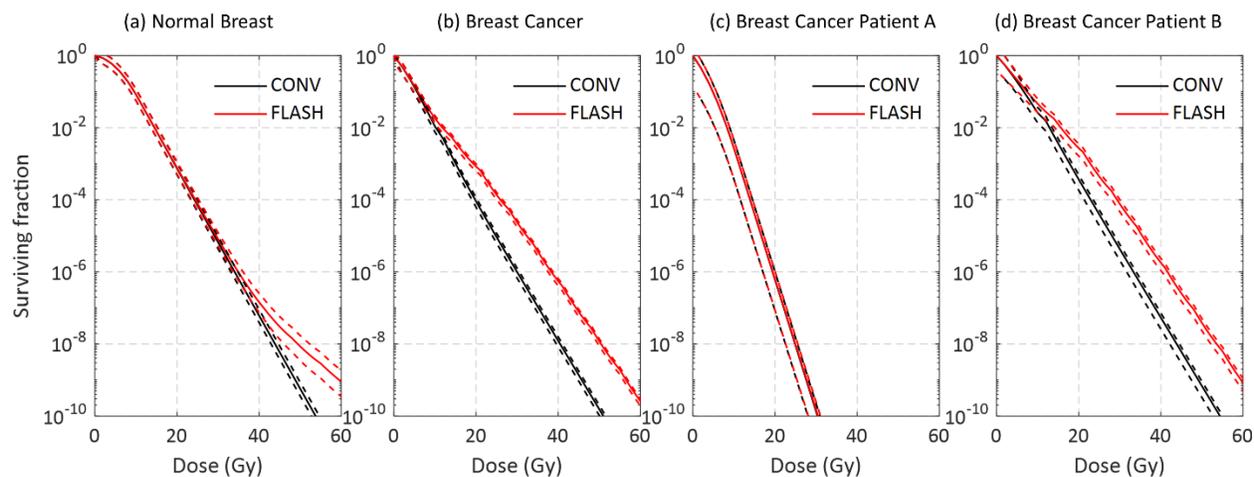

Figure 6. The surviving fractions of aggregate (a) normal and (b) tumor breast tissue, as well as (c-d) patient specific breast tumor following CONV or FLASH irradiation. Calculated with $g$ = 0.45 μM/Gy (0.36 mmHg/Gy) and $k$ = 3.8 mmHg. The dashed lines indicate the SF calculated with the 95% confidence intervals of pO$_2$ profiles in Figure 3.

The ratio of the FLASH to the CONV dose to achieve the same surviving fraction (Figure 6) is plotted in Figure 7. For aggregate breast tumors, and patient A and B breast tumors, FLASH hypoxia induction increases the dose of FLASH irradiation to achieve the same SF as 10 Gy CONV irradiation by factors of approximately 1.04, 1.0 and 1.12 respectively.

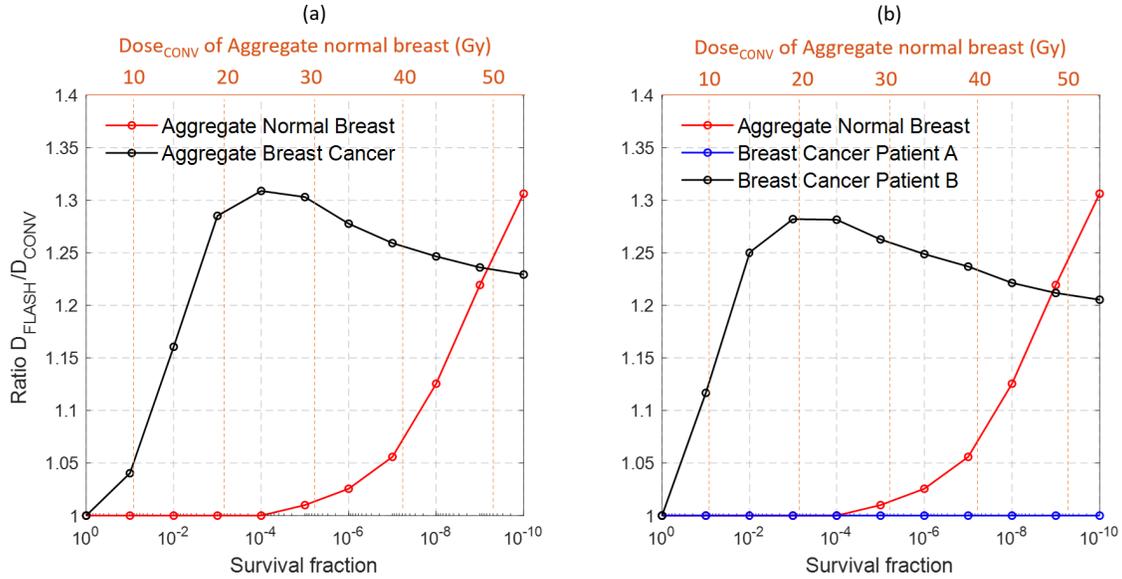

Figure 7. The ratio of the FLASH dose to the CONV dose to achieve the same surviving fraction is plotted as a function of dose. Panel (a) compares aggregate normal breast to aggregate breast tumor. Panel (b) compares normal breast to individual breast tumors of patient A and B. The upper X-axis indicate the CONV dose to reach the indicated SF in aggregate normal breast.

## 4 Discussion

The mechanism of the FLASH effect remains unclear, and previous studies indicate that the radical-radical recombination [11], tissue redox metabolism [47], altered immune/inflammatory response [48,49], and other mechanisms could play a significant role. Our results show that an evaluation of oxygen depletion as a potential mechanism of tissue sparing by FLASH irradiation requires examination of complete tissue $pO_2$ profiles. Predictions or assessments of response based on tissue median or average $pO_2$ can be significantly misleading. This pertains to CONV and especially FLASH irradiation.

Oxygen is a powerful modifier of cell and tissue response to radiation in the oxygen concentration range of 0-15 mmHg and especially in the 0-7 mmHg range. Our analysis shows that median or average $pO_2$ values do not reveal the fraction or the distribution of cells over that $pO_2$ range. For normal human brain with a median $pO_2$ of 25 mmHg, it may be assumed that it is more poorly oxygenated than human breast cancer which has a median $pO_2$ of 30 mmHg. However as seen in Figure 2, panels a and c, approximately 30% of breast tumor $pO_2$ values lie in the 0–10 mmHg $pO_2$ range vs. 11% in normal brain. FLASH significantly alters the response of both tissues, a finding not expected based on median tissue $pO_2$ values. Similarly, based on a median $pO_2$ value of 50 mmHg, the predicted response of subcutis to CONV and FLASH irradiation does not differ, but based on analysis of the tissue's complete profile, FLASH induces substantial radioprotection at doses above approximately 15 Gy.

We examined the impact of different model parameter values on cell response to irradiation. Varying the value of $k$ suggests that for aggregate or average breast tumor and normal tissue, an equal increase in the value of $k$ in both tissues decreases the dose at which hypoxic radioprotection is observed. This is especially true in tumor tissue due to its lower $pO_2$, (Figure 4b versus 4a). Studies have shown that the concentration of the principal cellular thiol, glutathione, in tumor cells, is higher when the cells are cultured in vitro, than in tumors formed from the same tumor cells, although the thiol concentration of tumors may not significantly differ from the thiol concentration of at least some normal tissues [22]. As pertains to $k$, an equal increase in $g$ would give rise to protection of both tissues at lower radiation doses (Figure 5). The very limited in vivo data pertaining to FLASH oxygen depletion in tumor versus normal tissue suggests that oxygen depletion in normal tissue may be greater than in tumors. As previously noted, Cao et al. reported that the $g$ value could be more than two times-higher in normal tissue than in tumor tissue [50].

In this study, the $pO_2$ profiles of human normal breast and breast tumor were examined because of the substantial $pO_2$ data available for both tissues. We also included brain and subcutis profiles of $pO_2$ and found that other human tumors and normal tissue $pO_2$ profiles can differ substantially from those observed in breast tissue. Similarly, the $pO_2$ profiles of rodent tumors and various rodent normal tissues may substantially differ from human breast tumor and normal tissue, although such rodent profiles have not been reported. Therefore, it cannot be concluded that the results presented here are in agreement with or differ from studies performed in rodents. However, the principles may be expected to apply to both. For example, for severely hypoxic tumors, as seen in a murine sarcoma tumor [51], further depleting oxygen by FLASH will not appreciably affect response. Similarly, analyses of normal human brain response to FLASH radiation might be expected to apply to normal rodent brain tissue if the $pO_2$ spectrums are similar. Although not seen in the normal tissue pO2 profiles presented here, the possible presence of a very small population of hypoxic normal tissue stem cells on the threshold of radiobiologic hypoxia (5-10 mmHg), as previously reported [6,52,53], cannot be excluded, as the Eppendorf system does not resolve pO$_2$ on a cell-by-cell basis. If present, FLASH oxygen depletion would likely contribute to their survival and possibly the repair of radiation injury. The immediate clinical applicability of our study is limited by the absence of readily accessible complete pO$_2$ profiles of human tumors and at-risk normal tissues. The development of non-invasive assays capable of resolving <1- 5% frequency pO$_2$ values of less than 10 mmHg, if present, on a patient by patient basis, would facilitate the choice of FLASH VS. conventional dose-rate irradiation.

The oxygen depletion model and its consequent effects on cell sensitivity developed in this study is based on radiolytic OH• production. OH• radical depletion is assumed to occur via the oxidation of biomolecule binding sites with a molar concentration in the intracellular environment, resulting in the production of biomolecular radicals R• (including DNA•) in approximately 10$^{-9}$s. The micromolar concentration of dissolved oxygen is depleted via oxidation of R• and DNA•. DNA• is either repaired by thiol or fixed by oxygen, and the oxygen depletion is a manifestation of oxygen enhancement/fixation of radiation damage. At a sufficiently large radiation dose, the oxygen concentration approaches and then decreases below the value of $k$, and thiol repair of damage predominates over the oxygen fixation. Therefore, our model assumed a dependent change of radiosensitivity parameters during the oxygen depletion of FLASH irradiation.  It should be noted, the model developed in this study does not assume or require the depletion of oxygen during a 3 ns pulse of radiation.  As noted above, the fate of the oxygen consuming DNA• is determined by the competition between oxygen which reacts with and fixes the DNA damage, and thiols which repair the DNA•.  In experimental studies in bacteria, Howard-Flanders and Moore showed that the lifespan of the oxygen consuming carbon centered radicals R• (such as DNA•) was less than 10 ms. Based on the diffusion coefficient of oxygen, the authors estimated the lifetime of R• to be approximately 10$^{-4}$ s in the absence of oxygen, and 10$^{-6}$ s in the presence of oxygen[54].

## 5 Conclusion

This study developed a method for analyzing oxygen depletion during extremely high dose-rate radiation and its consequent effect on cell and tissue response. Our method provides a framework that can be used to estimate which normal tissue and tumor circumstances may benefit from FLASH therapy. The results exposed the inadequacy of using median $pO_2$ for the prediction of tissue response to radiation due to oxygen depletion. Based on the complete $pO_2$ profile of normal human breast and tumor, in the majority of breast cases, CONV irradiation is superior to FLASH irradiation. The results obtained do not pertain to other rodent and human tissues' response to FLASH irradiation. Aggregate $pO_2$ profiles of normal breast, subcutis and brain reveal significant $pO_2$ profile differences, which will influence oxygen depletion mediated differences in response to FLASH irradiation.

## Reference

1. Favaudon V, Caplier L, Monceau V, et al. Ultrahigh dose-rate FLASH irradiation increases the differential response between normal and tumor tissue in mice [published online ahead of print 2014/07/18]. *Sci Transl Med.* 2014;6(245):245ra293.

2. Diffenderfer ES, Verginadis, II, Kim MM, et al. Design, Implementation, and in Vivo Validation of a Novel Proton FLASH Radiation Therapy System [published online ahead of print 2020/01/14]. *Int J Radiat Oncol Biol Phys.* 2020;106(2):440-448.

3. Levy K, Natarajan S, Wang J, et al. Abdominal FLASH irradiation reduces radiation-induced gastrointestinal toxicity for the treatment of ovarian cancer in mice [published online ahead of print 2020/12/12]. *Sci Rep.* 2020;10(1):21600.

4. Adrian G, Konradsson E, Lempart M, Back S, Ceberg C, Petersson K. The FLASH effect depends on oxygen concentration [published online ahead of print 2019/12/12]. *Br J Radiol.* 2020;93(1106):20190702.

5. Wardman P. Radiotherapy Using High-Intensity Pulsed Radiation Beams (FLASH): A Radiation-Chemical Perspective. *Radiation Research.* 2020;194(6):607-617.

6. Pratx G, Kapp DS. Ultra-High-Dose-Rate FLASH Irradiation May Spare Hypoxic Stem Cell Niches in Normal Tissues. *Int J Radiat Oncol.* 2019;105(1):190-192.

7. Spitz DR, Buettner GR, Petronek MS, et al. An integrated physico-chemical approach for explaining the differential impact of FLASH versus conventional dose rate irradiation on cancer and normal tissue responses [published online ahead of print 2019/04/24]. *Radiother Oncol.* 2019;139:23-27.

8. Wilson JD, Hammond EM, Higgins GS, Petersson K. Ultra-High Dose Rate (FLASH) Radiotherapy: Silver Bullet or Fool's Gold? *Frontiers in Oncology.* 2020;9:1563.

9.      Jin JY, Gu A, Wang W, Oleinick NL, Machtay M, Spring Kong FM. Ultra-high dose rate effect on circulating immune cells: A potential mechanism for FLASH effect? [published online ahead of print 2020/05/11]. *Radiother Oncol.* 2020;149:55-62.

10.     Petersson K, Adrian G, Butterworth K, McMahon SJ. A Quantitative Analysis of the Role of Oxygen Tension in FLASH Radiation Therapy [published online ahead of print 2020/03/08]. *Int J Radiat Oncol Biol Phys.* 2020;107(3):539-547.

11.     Labarbe R, Hotoiu L, Barbier J, Favaudon V. A physicochemical model of reaction kinetics supports peroxyl radical recombination as the main determinant of the FLASH effect. *Radiotherapy and Oncology.* 2020;153:303-310.

12.     Zhou G. Mechanisms underlying FLASH radiotherapy, a novel way to enlarge the differential responses to ionizing radiation between normal and tumor tissues. *Radiation Medicine and Protection.* 2020;1(1):35-40.

13.     Venkatesulu BP, Sharma A, Pollard-Larkin JM, et al. Ultra high dose rate (35 Gy/sec) radiation does not spare the normal tissue in cardiac and splenic models of lymphopenia and gastrointestinal syndrome [published online ahead of print 2019/11/22]. *Sci Rep.* 2019;9(1):17180.

14.     Smyth LM, Donoghue JF, Ventura JA, et al. Comparative toxicity of synchrotron and conventional radiation therapy based on total and partial body irradiation in a murine model. *Scientific reports.* 2018;8(1):1-11.

15.     Adrian G, Konradsson E, Beyer S, et al. Cancer cells can exhibit a sparing FLASH effect at low doses under normoxic in vitro-conditions. *Frontiers in oncology.* 2021.2890.

16.     Vozenin MC, De Fornel P, Petersson K, et al. The Advantage of FLASH Radiotherapy Confirmed in Mini-pig and Cat-cancer Patients [published online ahead of print 2018/06/08]. *Clin Cancer Res.* 2019;25(1):35-42.

17.     Field SB, Bewley DK. Effects of dose-rate on the radiation response of rat skin [published online ahead of print 1974/09/01]. *Int J Radiat Biol Relat Stud Phys Chem Med.* 1974;26(3):259-267.

18.     Soto LA, Casey KM, Wang J, et al. FLASH Irradiation Results in Reduced Severe Skin Toxicity Compared to Conventional-Dose-Rate Irradiation [published online ahead of print 2020/08/28]. *Radiat Res.* 2020;194(6):618-624.

19.     Bourhis J, Sozzi WJ, Jorge PG, et al. Treatment of a first patient with FLASH-radiotherapy [published online ahead of print 2019/07/16]. *Radiother Oncol.* 2019;139:18-22.

20.     Bedogni B, Powell MB. Skin hypoxia: a promoting environmental factor in melanomagenesis [published online ahead of print 2006/06/09]. *Cell Cycle.* 2006;5(12):1258-1261.

21.     Alexander P. Division of Biophysics: On the Mode of Action of Some Treatments That Influence the Radiation Sensitivity of Cells. *Transactions of the New York Academy of Sciences.* 1962;24(8 Series II):966-978.

22.   Koch C. The Mechanism of radioprotection by non-protein sulfhydryls: Cysteine, Glutathione, Cysteamine. In: "Radioprotectors: chemical, biological, and clinical perspectives" by E.A. Bump and K. Malaker. CRC press; 1988.

23.   Bump EA, Cerce BA, Al-Sarraf R, Pierce SM, Koch CJ. Radioprotection of DNA in isolated nuclei by naturally occurring thiols at intermediate oxygen tension. *Radiation research.* 1992;132(1):94-104.

24.   Dewey DL. The x-ray sensitivity of Serratia marcescens [published online ahead of print 1963/05/01]. *Radiat Res.* 1963;19(1):64-87.

25.   Dewey DL, Boag JW. Modification of the oxygen effect when bacteria are given large pulses of radiation [published online ahead of print 1959/05/23]. *Nature.* 1959;183(4673):1450-1451.

26.   Weiss H, Epp E, Heslin J, Ling C, Santomasso A. Oxygen depletion in cells irradiated at ultra-high dose-rates and at conventional dose-rates. *International Journal of Radiation Biology and Related Studies in Physics, Chemistry and Medicine.* 1974;26(1):17-29.

27.   Nias AH, Swallow AJ, Keene JP, Hodgson BW. Effects of pulses of radiation on the survival of mammalian cells [published online ahead of print 1969/07/01]. *Br J Radiol.* 1969;42(499):553.

28.   Epp ER, Weiss H, Djordjevic B, Santomasso A. The Radiosensitivity of Cultured Mammalian Cells Exposed to Single High Intensity Pulses of Electrons in Various Concentrations of Oxygen. *Radiation Research.* 1972;52(2):324-332.

29.   Suit HD, Maeda M. Hyperbaric oxygen and radiobiology of a C3H mouse mammary carcinoma [published online ahead of print 1967/10/01]. *J Natl Cancer Inst.* 1967;39(4):639-652.

30.   Khan S, Bassenne M, Wang J, et al. Multicellular Spheroids as In Vitro Models of Oxygen Depletion During FLASH Irradiation [published online ahead of print 2021/02/06]. *Int J Radiat Oncol Biol Phys.* 2021;110(3):833-844.

31.   Wright EA, Bewley DK. Whole-body radioprotection of mice to 8-Mev electrons produced by breathing nitrogen for brief periods. With a note on the dosimetry of whole-body irradiation of mice with fast electrons [published online ahead of print 1960/11/01]. *Radiat Res.* 1960;13(5):649-656.

32.   Wright EA, Batchelor AL. The change in the radiosensitivity of the intact mouse thymus produced by breathing nitrogen; with a note on the dosimetry for whole body fast electron irradiation [published online ahead of print 1959/03/01]. *Br J Radiol.* 1959;32(375):168-173.

33.   Beuve M, Alphonse G, Maalouf M, et al. Radiobiologic parameters and local effect model predictions for head-and-neck squamous cell carcinomas exposed to high linear energy transfer ions. *International Journal of Radiation Oncology* Biology* Physics.* 2008;71(2):635-642.

34.   Astrahan M. Some implications of linear‐quadratic‐linear radiation dose‐response with regard to hypofractionation. *Medical physics.* 2008;35(9):4161-4172.

35. Wouters BG, Brown JM. Cells at intermediate oxygen levels can be more important than the" hypoxic fraction" in determining tumor response to fractionated radiotherapy. *Radiation research.* 1997;147(5):541-550.

36. Alper T, Howard-Flanders P. Role of oxygen in modifying the radiosensitivity of E. coli B [published online ahead of print 1956/11/03]. *Nature.* 1956;178(4540):978-979.

37. Ling CC, Michaels HB, Epp ER, Peterson EC. Oxygen Diffusion into Mammalian Cells Following Ultrahigh Dose Rate Irradiation and Lifetime Estimates of Oxygen-Sensitive Species. *Radiation Research.* 1978;76(3):522-532.

38. Michaels HB, Epp ER, Ling CC, Peterson EC. Oxygen Sensitization of CHO Cells at Ultrahigh Dose Rates: Prelude to Oxygen Diffusion Studies. *Radiation Research.* 1978;76(3):510-521.

39. Vaupel P, Schlenger K, Knoop C, Hockel M. Oxygenation of human tumors: evaluation of tissue oxygen distribution in breast cancers by computerized O2 tension measurements [published online ahead of print 1991/06/15]. *Cancer Res.* 1991;51(12):3316-3322.

40. Vaupel P, Kallinowski F, Okunieff P. Blood flow, oxygen and nutrient supply, and metabolic microenvironment of human tumors: a review. *Cancer research.* 1989;49(23):6449-6465.

41. Zhou S, Zheng D, Fan Q, et al. Minimum Dose Rate Estimation for Pulsed FLASH Radiotherapy: A Dimensional Analysis. *Medical Physics.* 2020.

42. Gould MN, Howard SP. Radiosensitivity and PLDR in primary cultures of human normal and malignant mammary and prostate cells [published online ahead of print 1989/11/01]. *Int J Radiat Biol.* 1989;56(5):561-565.

43. Kehwar TS. Analytical approach to estimate normal tissue complication probability using best fit of normal tissue tolerance doses into the NTCP equation of the linear quadratic model [published online ahead of print 2007/11/14]. *J Cancer Res Ther.* 2005;1(3):168-179.

44. Bergsten P, Amitai G, Kehrl J, Dhariwal KR, Klein HG, Levine M. Millimolar concentrations of ascorbic acid in purified human mononuclear leukocytes. Depletion and reaccumulation. *Journal of Biological Chemistry.* 1990;265(5):2584-2587.

45. Clopper CJ, Pearson ES. The use of confidence or fiducial limits illustrated in the case of the binomial. *Biometrika.* 1934;26(4):404-413.

46. Dahlberg W, Little J, Fletcher J, Suit H, Okunieff P. Radiosensitivity in vitro of human soft tissue sarcoma cell lines and skin fibroblasts derived from the same patients. *International journal of radiation biology.* 1993;63(2):191-198.

47. Montay-Gruel P, Acharya MM, Petersson K, et al. Long-term neurocognitive benefits of FLASH radiotherapy driven by reduced reactive oxygen species [published online ahead of print 2019/05/18]. *Proc Natl Acad Sci U S A.* 2019;116(22):10943-10951.

48. Eggold JT, Chow S, Melemenidis S, et al. Abdominopelvic FLASH Irradiation Improves PD-1 Immune Checkpoint Inhibition in Preclinical Models of Ovarian Cancer [published online ahead of print 2021/12/07]. *Mol Cancer Ther.* 2022;21(2):371-381.

49. Zhu H, Xie D, Yang Y, et al. Radioprotective effect of X-ray abdominal FLASH irradiation: Adaptation to oxidative damage and inflammatory response may be benefiting factors [published online ahead of print 2022/04/23]. *Med Phys.* 2022;49(7):4812-4822.

50. Cao X, Zhang R, Esipova TV, et al. Quantification of Oxygen Depletion During FLASH Irradiation In Vitro and In Vivo [published online ahead of print 2021/04/13]. *Int J Radiat Oncol Biol Phys.* 2021;111(1):240-248.

51. Collingridge DR, Young WK, Vojnovic B, et al. Measurement of Tumor Oxygenation: A Comparison between Polarographic Needle Electrodes and a Time-Resolved Luminescence-Based Optical Sensor. *Radiation Research.* 1997;147(3):329-334.

52. Mohyeldin A, Garzon-Muvdi T, Quinones-Hinojosa A. Oxygen in stem cell biology: a critical component of the stem cell niche [published online ahead of print 2010/08/05]. *Cell Stem Cell.* 2010;7(2):150-161.

53. Pratx G, Kapp DS. A computational model of radiolytic oxygen depletion during FLASH irradiation and its effect on the oxygen enhancement ratio [published online ahead of print 2019/08/01]. *Phys Med Biol.* 2019;64(18):185005.

54. Howard-Flanders P, Moore D. The time interval after pulsed irradiation within which injury to bacteria can be modified by dissolved oxygen: I. A search for an effect of oxygen 0.02 second after pulsed Irradiation. *Radiation Research.* 1958;9(4):422-437.

**Supplementary Material**

## 1. Summary of previously reported values of g

Weiss *et al*. measured the value of *g* by irradiating an oxygen-equilibrated bacterial cell suspension in a sealed vessel and reported $g = 0.58 \pm 0.1$ µM/Gy under conventional dose rate and $g = 0.26 \pm 0.05$ µM/Gy at ultra-high dose rate thin layer conditions i.e. bacteria coated with a film of culture medium [1]. Michaels measured the value of *g* in stirred aqueous solutions of CHO cells in sealed glass vessels and reported $g = 0.44$ µM/Gy [2], similar to the value of *g* evaluated in a thin layer technique $g = 0.48$ µM/Gy [3]. Epp *et al*. reported $g = 0.61 - 0.71$ µM/Gy in HeLa cells [4]. Nias *et al*. reported $g = 0.65$ µM/Gy in HeLa cells using 1 µs pulsed electrons [5]. Boscolo *et al*. reported $g = 0.33$ µM/Gy for 1 MeV electrons by Monte Carlo (MC) simulation [6]. Lai *et al*. reported $g = 0.19 - 0.22$ µM/Gy for 4.5 keV electrons at a dose rate of $10^6 - 10^8$ Gy/s by MC simulation [7], and Zhu *et al*. obtained $g = 0.38 - 0.43$ µM/Gy for 4.5 MeV electrons with MC simulation [8].

## 2. The method to calculate 95% confidence intervals of pO$_2$ distribution

The confidence interval is related to the number of measured foci in a particular pO$_2$ bin, and it is the confidence interval of a proportion. The upper and lower boundaries of the 95% confidence intervals are calculated according to the reference [9] with the following equation (1) and (2) respectively.

$$\text{upper boundary} = p + z \sqrt{\frac{p(1-p)}{n}} \qquad (1)$$

$$\text{lower boundary} = p - z \sqrt{\frac{p(1-p)}{n}} \qquad (2)$$

where *p* is the proportion value in a particular bin; *n* is the sample number (total number of measured foci); The number *z* defines the size of the confidence interval of a normal/gaussian distribution, if $z = 1$, then approximately 68% of the values lie within one standard deviation, and $z = 1.96$ for the 95% confidence interval.

## 3. Reproducing experimental data (HeLa cells)

The following figure shows the SF curves of HeLa cells obtained by Epp et al. [4] and reproduced with the method described in section 2.2 of the manuscript. It should be noted that the original data extracted from Epp et al. were uncorrected for cellular multiplicity ($N$), which was estimated average $N$ =2. Cellular multiplicity refers to the the number of cells per colony forming unit. The determination of SF curve inactivation parameters requires single cell suppensions with $N$ =1. Assuming the survival probability of a colony forming unit after a dose of $D$ is $m$, then the survival probability of a cell in a colony forming unit with a cellular multiplicity of $N$ is SF = 1- $(1-m)^{1/N}$. If uncorrected $N$ higher than 1 results in shallower curve slope and incorrect SF shape. To correct for the cellular multiplicity effect, the experimental data from Epp et al. were corrected with $SF_{corrected} = 1 - \sqrt{1 - SF_{Epp}}$ where $SF_{epp} =$ m. More comprehensive corrections requires knowledge of the discreet multiplicity (not just the average multiplicity, and the value of "m" at zero dose [10]. These data were not provided.

After data correction, the $\alpha$ and $\beta$ values were derived from the $N_2$ curve with $D_T$ = 12 Gy and used to reproduce the SF curve of other oxygen concentrations (with OER adjustments of $D_T$). The following figure shows that our reproduced dashed lines fit the experimental data in the high dose region. Data and fits to two additional oxygen concentrations < 0.91% $O_2$, is not plotted for clarity. In general, there was substantial scatter in the experimental data at SF values > 0.1 but the agreement between the predicted and observed SF values was strong at SF < 0.1. Figure 1 of the manuscript shows fits of additional data sets.

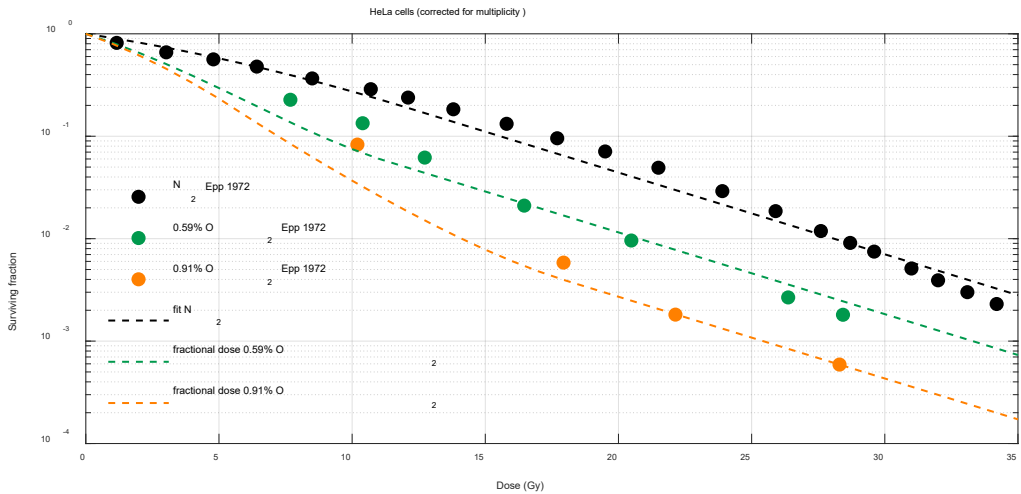

Figure S1. The SF curves of HeLa cells exposed to simulated FLASH irradiation along with the experimental data of Epp et al. [4]. $\alpha_{anoxic}$=0.09097, $\beta_{anoxic}$=0.003867, $D_T$=12 Gy, $g$=0.368 mmHg/Gy.

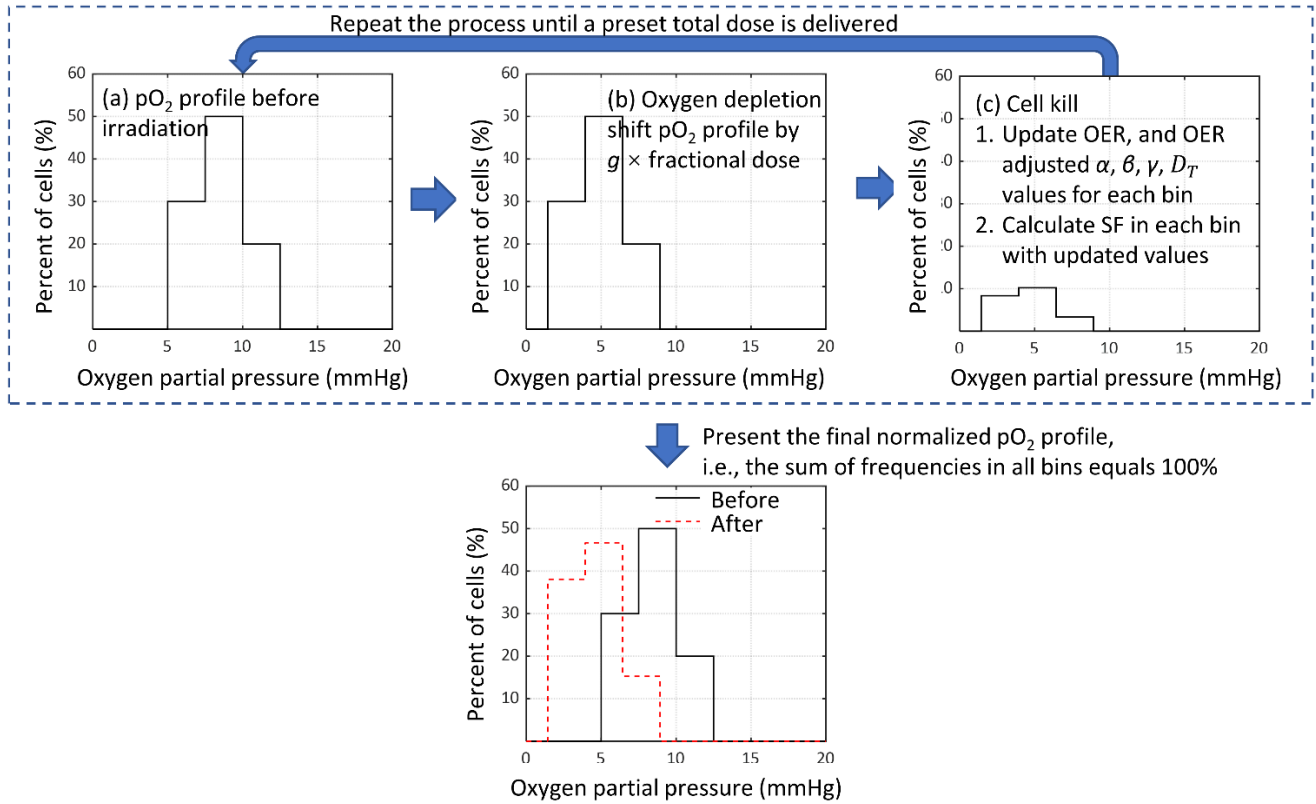

Figure S2. Graphic description of analysis method based on the full pO$_2$ profile. ① shift the pO$_2$ profile by $g \times$ fractional dose; ② calculate the surviving fraction in each bin using the adjusted $\alpha$, $\beta$, $\gamma$, $D_T$ values; ③ repeat the process until a preset total dose is delivered; ④ normalize the pO$_2$ profile so that the sum of frequencies in all bins equals 100%.

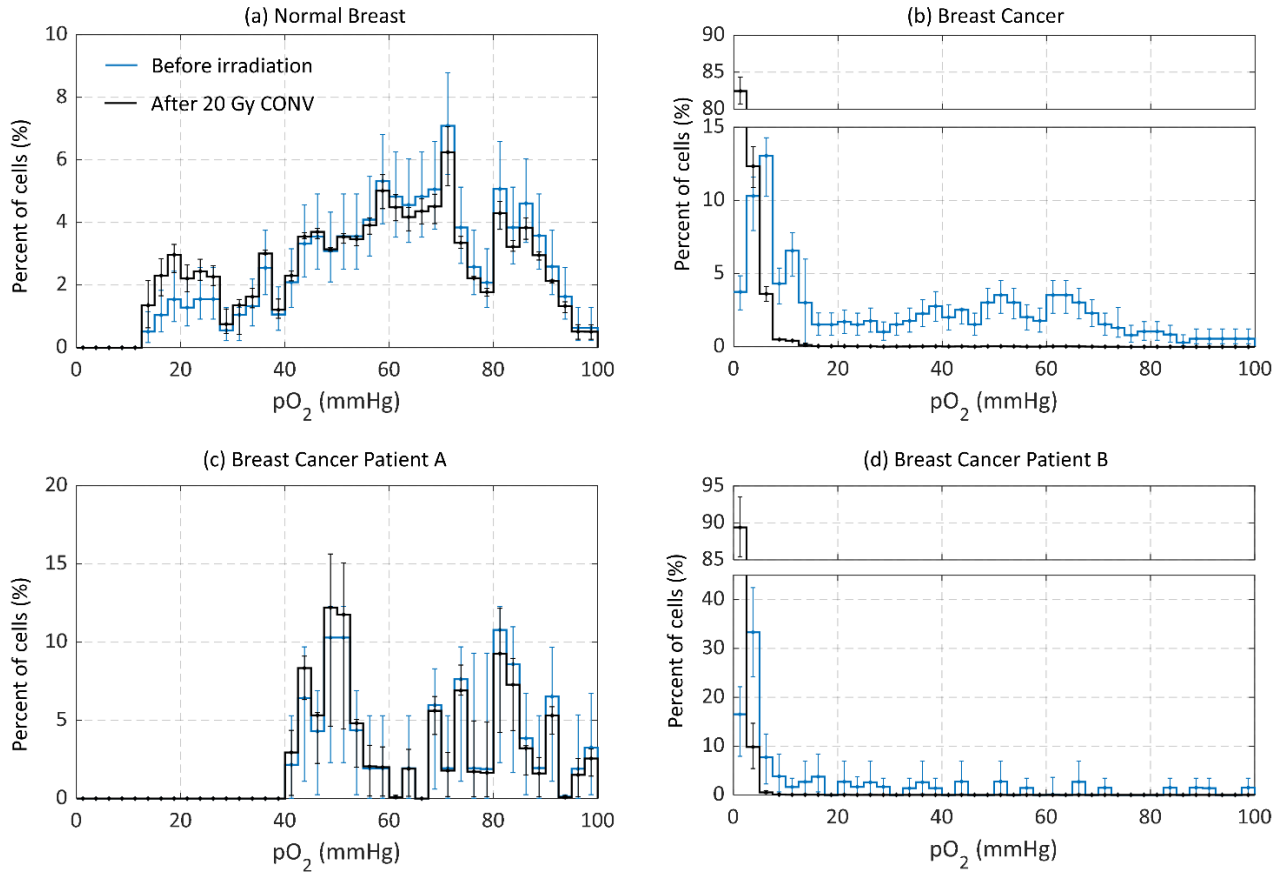

Figure S3. The pO$_2$ profile of (a) aggregate normal breast; (b) aggregate breast cancer; and (c-d) patient specific breast tumors before (blue lines The error bars in panels a - d are 95% confidence intervals. Calculated with g = 0.45 μM/Gy (0.36 mmHg/Gy).

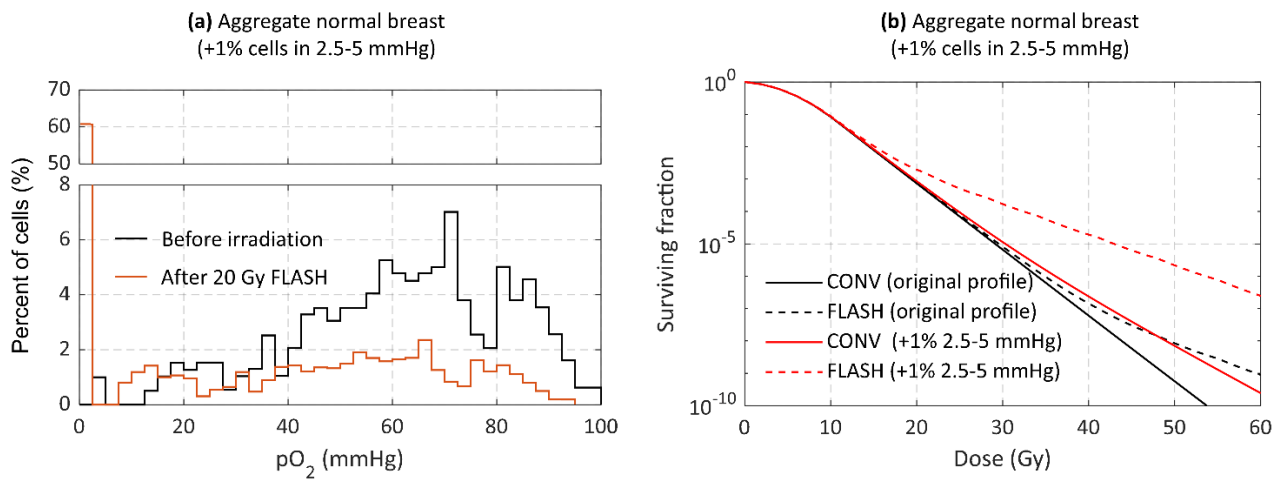

Figure S4. The impact of 1% cells on the threshold of hypoxia. (a) The pO$_2$ profile of aggregate normal breast with 1% tissue in the 2.5-5 mmHg bin. (b) The SF curves of original aggregate normal breast (black lines) and aggregate normal breast with 1% tissue in the 2.5-5 mmHg bin (red lines).

## Reference

1.  Weiss H, Epp E, Heslin J, Ling C, Santomasso A. Oxygen depletion in cells irradiated at ultra-high dose-rates and at conventional dose-rates *International Journal of Radiation Biology and Related Studies in Physics, Chemistry and Medicine.* 1974;26(1):17-29.

2.  Michaels HB. Oxygen depletion in irradiated aqueous solutions containing electron affinic hypoxic cell radiosensitizers *Int J Radiat Oncol Biol Phys.* 1986;12(7):1055-1058.

3.  Michaels HB, Epp ER, Ling CC, Peterson EC. Oxygen Sensitization of CHO Cells at Ultrahigh Dose Rates: Prelude to Oxygen Diffusion Studies *Radiation Research.* 1978;76(3):510-521.

4.  Epp ER, Weiss H, Djordjevic B, Santomasso A. The Radiosensitivity of Cultured Mammalian Cells Exposed to Single High Intensity Pulses of Electrons in Various Concentrations of Oxygen *Radiation Research.* 1972;52(2):324-332.

5.  Nias AH, Swallow AJ, Keene JP, Hodgson BW. Effects of pulses of radiation on the survival of mammalian cells *Br J Radiol.* 1969;42(499):553.

6.  Boscolo D, Scifoni E, Durante M, Kramer M, Fuss MC. May oxygen depletion explain the FLASH effect? A chemical track structure analysis *Radiother Oncol.* 2021;162:68-75.

7.  Lai Y, Jia X, Chi Y. Modeling the effect of oxygen on the chemical stage of water radiolysis using GPU-based microscopic Monte Carlo simulations, with an application in FLASH radiotherapy *Physics in Medicine & Biology.* 2020. doi: 10.1088/1361-6560/abc93b/meta.

8.  Zhu H, Li J, Deng X, Qiu R, Wu Z, Zhang H. Modeling of cellular response after FLASH irradiation: a quantitative analysis based on the radiolytic oxygen depletion hypothesis *Phys Med Biol.* 2021;66(18).

9.  Clopper CJ, Pearson ES. The use of confidence or fiducial limits illustrated in the case of the binomial *Biometrika.* 1934;26(4):404-413.

10. Gerweck L, Dullea R, Zaidi S, Budach W, Hartford A. Influence of experimental factors on intrinsic radiosensitivity assays at low doses of radiation: cell multiplicity *Radiation research.* 1994;138(3):361-366.